\documentclass[twocolumn,secnumarabic,amssymb, nobibnotes, aps]{revtex4-1}
\usepackage{graphicx}
\usepackage{dcolumn}
\usepackage{bm}

\begin{document}

\title{Zero-bias anomaly in a nanowire quantum dot coupled to superconductors}%

\author{Eduardo J. H. Lee$^{1}$}
\author{Xiaocheng Jiang$^{2}$}
\author{Ram\'{o}n Aguado$^{3}$}
\author{Georgios Katsaros$^{1}$}
\author{Charles M. Lieber$^{2}$}
\author{Silvano De Franceschi$^{1}$}
\email{silvano.defranceschi@cea.fr}

\affiliation{$^{1}$SPSMS, CEA-INAC/UJF-Grenoble 1, 17 rue des Martyrs, 38054 Grenoble Cedex 9, France}
\affiliation{$^{2}$Harvard University, Department of Chemistry and Chemical Biology, Cambridge, MA, 02138, USA}
\affiliation{$^{3}$Instituto de Ciencia de Materiales de Madrid, ICMM-CSIC Cantoblanco, 28049 Madrid, Spain}

\date{\today}%

\begin{abstract}
We studied the low-energy states of spin-1/2 quantum dots defined in InAs/InP nanowires and coupled to aluminium superconducting leads. By varying the superconducting gap, $\Delta$, with a magnetic field, $B$, we investigated the transition from strong coupling, $\Delta \ll T_K$, to weak coupling, $\Delta \gg T_K$, where $T_{K}$ is the Kondo temperature. Below the critical field, we observe a persisting zero-bias Kondo resonance that vanishes only for low $B$ or higher temperatures, leaving the room to more robust sub-gap structures at bias voltages between $\Delta$ and $2\Delta$. For strong and approximately symmetric tunnel couplings, a Josephson supercurrent is observed in addition to the Kondo peak. We ascribe the coexistence of a Kondo resonance and a superconducting gap to a significant density of intra-gap quasiparticle states, and the finite-bias sub-gap structures to tunneling through Shiba states. Our results, supported by numerical calculations, own relevance also in relation to tunnel-spectroscopy experiments aiming at the observation of Majorana fermions in hybrid nanostructures.

\begin{description}
\item[PACS numbers]
72.15.Qm, 73.21.La, 73.63.Kv, 74.45.+c
\end{description}
\end{abstract}

\pacs{Valid PACS appear here}

\maketitle

Hybrid devices which couple superconducting (\textit{S}) electrical leads to low-dimensional semiconductors have received great attention due to their fascinating underlying physics\cite{ReviewDF}. Further interest in this field has been generated by recent theoretical predictions on the existence of Majorana fermions at the edges of one-dimensional semiconductor nanowires (NWs) with strong spin-orbit interaction connected to \textit{S} electrodes \cite{DasSarma}. Zero-bias conductance peaks meeting some of the expected characteristic signatures of Majorana physics were recently reported in hybrid devices based on InSb \cite{MajoDelft, MajoXu} and InAs NWs \cite{Weizmann}. In the past years, quantum dots (QDs) coupled to superconducting leads have been widely explored as tunable Josephson junctions\cite{sdf2006,Cleuziou}, or as building blocks of Cooper-pair splitters \cite{splitter01,splitter02,Dascooper}. Hybrid superconductor-QD devices also constitute versatile platforms for studying fundamental issues, such as the physics of the Andreev bound states (ABS) \cite{Pillet,Deacon2010,Simon} or the interplay between the Kondo effect and the superconducting proximity effect\cite{Buitelaar,Siano,Choi,Oguri,Buizert,Eichler,Jesp,Kanai,Kasper,Eichler02,stm}.

The Kondo effect usually stems from the antiferromagnetic coupling of a localized electron spin and a Fermi sea of conduction electrons. Below a characteristic temperature $T_K$, the so-called Kondo temperature, a many-body spin-singlet state is formed, leading to the partial or complete screening of the local magnetic moment. This phenomenon, discovered in metals containing diluted magnetic impurities, is now routinely found in individual QDs with a spin-degenerate ground state, e.g. QDs hosting an odd number of electrons. The Kondo effect manifests itself as a zero-bias conductance peak whose width is proportional to $T_{K}$. In \textit{S}-QD-\textit{S} devices, the quasiparticle density of states (DOS) around the Fermi level ($E_{F}$) of the leads vanishes due to the opening of the superconducting gap ($\Delta$). This lack of quasiparticles precludes Kondo screening. 

The competition between the Kondo effect and superconductivity is governed by the corresponding energy scales, $k_{B}T_{K}$ and $\Delta$. While no Kondo screening occurs for $k_{B}T_{K} \ll \Delta$ (weak coupling), a Kondo singlet is expected to form for $k_{B}T_{K} \gg \Delta$ (strong coupling) at the expense of the breaking of Cooper pairs at the Fermi level  \cite{Buizert, Glazman}. A quantum phase transition is predicted to take place at $k_{B}T_{K} \approx \Delta$ \cite{Siano,Choi,Oguri}. Experimental signatures of this exotic crossover have been investigated both in the Josephson supercurrent regime \cite{Kasper,Eichler02,Kanai} and in the dissipative sub-gap transport regime \cite{Buizert,Eichler,Jesp}. Yet a full understanding of these experimental findings is still lacking. 
In this Letter, we report an experimental study on \textit{S}-QD-\textit{S} devices where the relative strength between Kondo and superconducting pairing correlations is tuned by means of a magnetic field, $B$, acting on $\Delta$. The transition from strong to weak coupling is continuously achieved by sweeping $B$ from above the critical field, $B_c$, where $\Delta = 0$, to zero field, where $\Delta$ attains its maximum value, $\Delta_{0}$, exceeding $k_B T_K$. 

The \textit{S}-QD-\textit{S} devices were fabricated from individual InAs/InP core/shell NWs grown by thermal evaporation (total diameter $\approx 30$ nm). The InP shell (thickness $\approx 2$ nm) acts as a confinement barrier resulting in an enhanced mobility of the one-dimensional electron gas in the InAs core \cite{Xiaocheng}. After growth, the NWs were deposited onto a degenerately doped, p-type Si substrate (used as a back gate), covered by a 300-nm-thick thermal oxide. Device fabrication was accomplished by e-beam lithography, Ar$^{+}$ bombardment (to remove native oxides), metal evaporation, and lift-off. Source and drain contacts consisted of Ti (2.5 nm)/Al (45 nm) bilayers with a lateral separation of approximately 200 nm, and a superconducting critical temperature of $\approx 1$ K. Transport measurements were performed in a He$_{3}$-He$_{4}$ dilution refrigerator with a base temperature of 15 mK. 
 
\begin{figure}
\includegraphics[width=86mm]{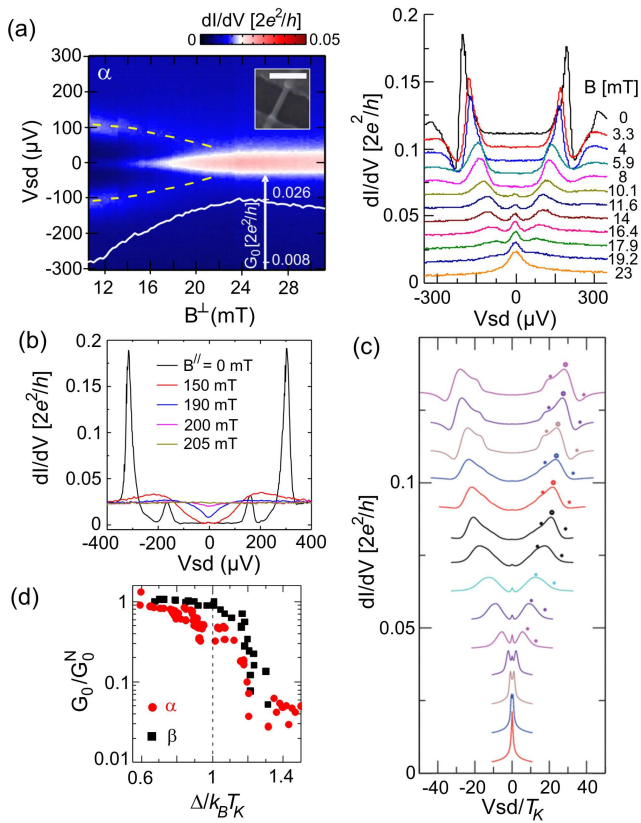}
\caption{\label{fig:epsart} (a) Left panel: Color plot of $dI/dV$ vs $(B^{\perp}, V_{SD})$ measured at the center of diamond $\alpha$. The dashed lines highlight the emergence of finite-bias peaks related to the opening of a superconducting gap. The superimposed line trace shows the  $B^{\perp}$-dependence of the linear conductance. Right panel: $dI/dV(V_{SD})$ traces taken at different $B^{\perp}$ values. The inset shows a scanning electron micrograph of a typical device (scale bar: 200 nm). (b) $dI/dV(V_{SD})$ traces measured in an even diamond revealing the Dynes-like DOS of leads. (c) Numerical calculations. The smaller solid dots denote the position of the $\Delta$ and 2$\Delta$ peaks, whereas the open dots highlight the position of the Shiba bound state peaks. (d) Linear conductance ($G_{0}$) normalized to the normal-state value ($G^{N}_{0}$) and plotted as a function of $\Delta\slash k_{B}T_{K}$ for diamonds $\alpha$ (red dots) and $\beta$ (black squares).}
\end{figure}

At low temperature, electron transport is dominated by Coulomb blockade with the NW channel behaving as a single QD. Charge stability measurements (i.e. differential conductance, $dI/dV$, as a function of source-drain bias, $V_{sd}$, and back-gate voltage, $V_G$) were performed to identify Kondo resonances in Coulomb diamonds with a spin-1/2 ground state. These measurements were taken at 15 mK with the leads in the normal state (superconductivity was suppressed by means of a magnetic field $B^{\perp}$ = 70 mT perpendicular to the substrate and exceeding the perpendicular critical field $B^{\perp}_{c}$). In each Kondo diamond, $T_K$ was measured from the half width at half maximum (HWHM) of the zero-bias $dI/dV$ peak, while the tunnel coupling asymmetry was extracted from the peak height, i.e. the linear conductance $G$, according to the relation: $G\slash G_{0}=4\Gamma_{L}\Gamma_{R}\slash(\Gamma_{L}+\Gamma_{L})^{2}$, where $G_{0}=2e^{2}\slash h$ and $\Gamma_{L(R)}$ is the tunnel coupling to the left (right) lead. Here we present data corresponding to three Kondo diamonds labeled as $\alpha$, $\beta$ and $\gamma$, where: $T_{K,\alpha} \approx$ 0.56 K, $(\Gamma_{L}\slash\Gamma_{R})_{\alpha} \approx$ 6.6 $\times$ 10$^{-3}$, $T_{K,\beta} \approx$ 1 K, $(\Gamma_{L}\slash\Gamma_{R})_{\beta} \approx$ 0.44, and $T_{K,\gamma} \approx$ 0.71 K, $(\Gamma_{L}\slash\Gamma_{R})_{\gamma} \approx 7.6 \times$ 10$^{-3}$.

Figure 1a shows a $dI/dV(B^{\perp}, V_{sd})$ measurement taken at the center of Kondo diamond $\alpha$. The zero-bias Kondo peak is apparent above $B_{c}^{\perp} \approx$ 23 mT 
(we note that at such low fields the Zeeman splitting is much smaller than $k_B T_{K,\alpha}$, explaining the absence of a split Kondo peak). 
Surprisingly, reducing the field below $B_{c}^{\perp}$ does not lead to an abrupt suppression of the Kondo peak. 
Instead, the peak becomes progressively narrower and smaller, vanishing completely only below $B^{\perp} \approx$ 9 mT. 

We argue that the observed zero-bias peak is a manifestation of Kondo screening due to intra-gap quasiparticle states. To support this interpretation, we show in Fig. 1b a data set
taken in an adjacent diamond with even occupation (i.e. with no Kondo effect). The $dI/dV(V_{sd})$ traces shown correspond to different in-plane fields, $B^{\parallel}$, ranging from zero to just above the in-plane critical field $B_{c}^{\parallel} \approx 200$ mT. When lowering the fields from above to below $B_{c}^{\parallel}$, the sub-gap $dI/dV$ does not drop abruptly, supporting our hypothesis of a sizable quasiparticle DOS at $E_F$  (we note that, although the measurement of Fig. 1b refers to in-plane fields, a qualitatively similar behavior can be expected for perpendicular fields, see Supplemental Material). The development of a "soft" gap just below $B_{c}^{\parallel}$ is additionally marked by the absence of $dI/dV$ peaks characteristic of the BCS DOS singularities (a discussion of the origin of the "soft" gap is included in the Supplemental Material). Such peaks develop only at fields well below $B_{c}^{\parallel}$, becoming most pronounced at $B=0$. In this low-field limit, the sub-gap conductance simultaneously vanishes and first-order multiple-Andreev-reflection resonances emerge at $eV_{sd} \approx \pm \Delta$. 

To reproduce the observed sub-gap features and the coexistence of a Kondo peak and a superconducting gap, we calculated the $dI/dV$ of a QD modeled by an Anderson Hamiltonian including coupling to BCS-type superconducting reservoirs. We used the so-called non-crossing approximation, a fully non-perturbative theory that includes both thermal and quantum fluctuations, complemented with the Keldysh-Green's function method to take into account non-equilibrium effects at finite $V_{sd}$ (see Supplemental Material). 
In order to fit the experimental data, the DOS of the leads was modeled by a Dynes function:  
\begin{equation} 
N_{s}(E,\gamma,B)=Re[\frac{|E|+i\gamma(B)}{\sqrt{(|E|+i\gamma(B))^2-\Delta(B)^2}}],
\end{equation} where $\gamma(B)$ is a phenomenological broadening term \cite{gamma}. For small $\Delta(B)$, the $\gamma(B)$ term is particularly important leading to a finite quasiparticle DOS at $E_F$. By contrast, as $\Delta(B)$ increases (with decreasing $B$), the DOS of the superconducting leads approaches the ideal BCS profile.

Figure 1c shows a set of calculated $dI/dV(V_{sd})$ traces at different $B$. 
By adjusting the DOS parameters $\gamma(B)$ and $\Delta(B)$, these calculations clearly show a zero-bias peak persisting below the $B_{c}$, in agreement with the experimental data of Fig. 1a. Since this peak emerges only in the case of a finite $\gamma(B)$, we conclude that the finite DOS at the Fermi level is at the origin of the experimentally observed zero-bias anomaly. The narrowing of this Kondo anomaly with increasing $\Delta$ can be interpreted as a decreasing $T_K$ due to the shrinking quasiparticle DOS around the Fermi level (this aspect is more quantitatively discussed in the Supplemental Material). As $T_K$ approaches the electronic temperature, the peak height gets smaller leading to the disappearance of the zero-bias peak. 

Figure 2 shows a second data set taken in Kondo diamond $\beta$. In this case, the stronger and more symmetric coupling to the leads results in a higher $T_{K}$, and a larger peak conductance. Nevertheless, the field dependence (Fig. 2a) shows substantially the same behavior as in Fig. 1a, i.e. a zero-bias Kondo peak persisting below $B^{\perp}_{c}$, becoming progressively narrower with decreasing $B^{\perp}$, and vanishing below $B^{\perp} \approx$ 10 mT. Interestingly, a sharp $dI/dV$ resonance is found around $V_{sd}=0$ superimposed
to the (wider) zero-bias Kondo peak. This resonance persists throughout the entire field range in which the leads are superconducting. By performing current-bias measurements (inset of Fig. 2b), we were able to ascribe this sharp resonance to a Josephson supercurrent as high as 0.9 nA at $B = 0$. This finding reveals the possibility of a coexistence between the Josephson effect, linked to the superconducting nature of the leads, and a Kondo effect arising from the exchange coupling between the localized electron and intra-gap quasiparticle states. 

The peak heights of Kondo resonances $\alpha$ and $\beta$ appear to follow approximately the same dependence on $\Delta\slash k_{B}T_{K}$ (Fig. 1d), where $T_K$ refers to the Kondo temperature in the normal state. A similar scaling was reported earlier by Buizert et al. \cite{Buizert}, for a \textit{S}-QD-\textit{S} device fabricated from an InAs self-assembled QD using Ti/Al contacts. In that paper, it was speculated that when $k_{B}T_{K} \gg \Delta$, it becomes energetically favorable for Cooper pairs to split in order to screen the 
local spin and create a Kondo resonance at the Fermi level. The results presented here point at a different interpretation based on the presence of the already discussed intra-gap quasiparticle states, which become particularly important when $B$ approaches $B_c$. 
According to this interpretation, the apparent scaling in Fig. 1d is intimately related to a quasi-particle "poisoning" of the superconducting gap. 

Well below $B_{c}$, as the Kondo anomaly disappears, the $dI/dV(V_{sd})$ is dominated by a pair of peaks symmetrically positioned with respect to $V_{sd}=0$ (Fig. 1a). 
These peaks become most pronounced at $B=0$. 
Similar types of sub-gap structures (SGS) have been reported in earlier works and  were given different interpretations: a Kondo enhancement of the first order Andreev reflection process \cite{Jesp,Buizert}, or, in the case of asymmetrically coupled \textit{S}-QD-\textit{S} devices, a persisting Kondo resonance, created by the strongly coupled lead, which is probed by the BCS DOS of the second, weakly-coupled lead \cite{Eichler}. 
Some of the above interpretations \cite{Buizert,Eichler}  invoke Kondo correlations to explain the observed sub-gap structure, even though, as we have pointed out, these correlations get suppressed as $\Delta$ reaches its largest value at $B=0$. 

\begin{figure}
\includegraphics[width=86mm]{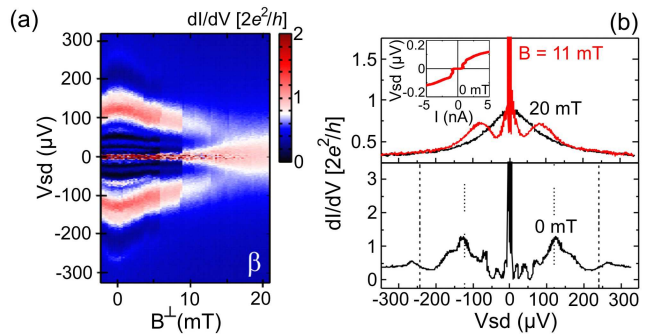}
\caption{\label{fig:epsart} (a) $B^{\perp}$-dependence measured in diamond $\beta$. (b) $dI/dV$ line profiles taken at different $B$. The inset is a voltage-current measurement carried out at $B = 0$, which reveals transport of a dissipationless supercurrent in the device.}
\end{figure}

More recently, finite-bias SGS were explained in terms of tunneling through Yu-Shiba-Rusinov states  \cite{Paaske2010, Paaske2011}. These intra-gap bound states, often referred to as Shiba states, were originally discussed in the case of magnetic impurities embedded in a superconductor \cite{Yu, Shiba, Rusinov,Balatsky}. They can be seen as ABS emerging as a result of the exchange coupling, $J$, between the impurity and the superconductor. As later confirmed by experiments based on scanning tunneling spectroscopy \cite{Yazdani}, Shiba states emerge as pairs of peaks in the local DOS symmetrically positioned at energies $\pm E_{B}$ relative to the Fermi level, where $E_{B}$ depends on $J$ and $\vert E_{B} \vert  < \Delta$.  

The zero-field $dI/dV(V_{sd})$ trace in the right panel of Fig. 1a exhibits a pair of $dI/dV$ peaks at $eV_{sd} \approx \pm 1.4\Delta$, followed by negative $dI/dV$ regions. Taking into account the strong asymmetry in the tunnel couplings, these features can be well explained in terms of a pair of Shiba levels with $E_{B} = 0.4 \Delta$, created by the strongly coupled S lead, and tunnel-probed by the weakly coupled S lead (see Fig. 3a). The observed $dI/dV$ peaks result from the alignment of these Shiba levels with the BCS gap-edge singularities. Precisely, one $dI/dV$ peak is due to the onset of electron tunneling from the Shiba level below $E_F$ to the empty quasi-particle band of the \textit{S} probe. The other peak is due to the onset of electron tunneling from the occupied quasi-particle band of the \textit{S} probe to the Shiba level above $E_F$. Increasing $\vert V_{sd} \vert$ beyond these resonance conditions leads to a reduced tunneling probability and hence a negative $dI/dV$.
The Shiba-related features observed if Fig. 1a are very well reproduced by the numerical results in Fig. 1c.   
In the case of relatively low contact asymmetry (lower panel of Fig. 2b), both \textit{S} leads interact with the QD spin resulting in a stronger $J$. We find a pair of $dI/dV$ peaks at $eV_{sd} \approx \pm\Delta$, which implies $E_{B} \approx 0$. In addition, we observe a rather complex set of smaller peaks most likely due to multiple Andreev reflection processes. 

\begin{figure}
\includegraphics[width=86mm]{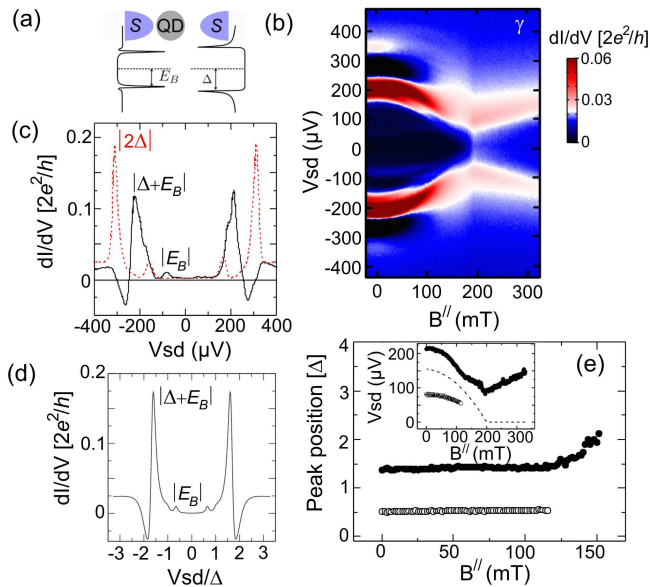}
\caption{\label{fig:epsart} (a) Schematics of the formation of Yu-Shiba-Rusinov bound states resulting from the interaction of the QD with the strongly-coupled lead. (b) $B^{\parallel}$-dependence of $dI/dV$ measured in diamond $\gamma$. (c) The solid line depicts the $dI/dV$ taken at $B = 0$. Two pairs of peaks are observed at $|eV_{sd}| = \Delta + E_{B}, E_{B}$. The dashed line shows the equivalent $dI\slash dV$ taken in an even diamond, for comparison. (d) Numerical calculation of the $dI/dV$ for $B$ = 0.
(e) Position of the intra-gap peaks as a function of $B$. The inset shows the data before performing the rescaling in units of $\Delta$. The dashed line displays the field-dependence of $\Delta$.}
\end{figure}

In the weak-coupling limit, the energy of the Shiba states is related to $\Delta$ through $E_{B} = \Delta(1-x)\slash (1+x)$, where $x = 3(\pi\nu_{F}J\slash 4)^2$ and $\nu_{F}$ is the Fermi velocity \cite{Paaske2011}. Since $J$ is presumably independent of $B$, the energy of the Shiba states should evolve proportionally to $\Delta(B)$ as $B$ is varied. We have verified this dependence with a measurement performed in diamond $\gamma$, where, as in diamond $\alpha$, tunnel couplings are strongly asymmetric. Differently from 
diamond $\alpha$, however, the low-energy transport is characterized by two pairs of intra-gap $dI/dV$ peaks, as shown by the zero-field curve in Fig. 3c. As in Fig. 1a, the most prominent peaks (at $eV_{sd} \sim \pm 1.4\Delta$) correspond to the alignment of the Shiba levels with the BCS coherence peaks of the weakly coupled contact, and they are consistently followed by negative $dI/dV$ dips. The weaker peaks at $eV_{sd} \approx \pm 0.53\Delta$ (in fact only one of them is clearly visible) can be interpreted as 'replicas' of the Shiba peaks. Such 'replicas' are expected when the Shiba levels line up with the Fermi level of the weakly coupled \textit{S} lead (for $eV_{sd} = \pm E_{B}$), provided a non negligible density of quasiparticles is present throughout its superconducting gap. This is apparent from the calculated $dI/dV(V_{sd})$ trace in Fig. 3d, which agrees fairly well with the experimental one. 

The Shiba peaks and their replicas shift towards $V_{sd}=0$ as $B^{\parallel}$ is increased. Their positions are plotted in the inset of Fig. 3e as solid and open dots, respectively.  For comparison, the $\Delta(B^{\parallel})$ dependence measured in a non-Kondo diamond is also plotted.  Up to $B^{\parallel} \sim$ 120 mT, the Shiba peaks and their replicas evolve proportionally to $\Delta(B^{\parallel})$ in agreement with the theoretical prediction. Above $B^{\parallel} \sim$ 120 mT, this behavior begins to be affected by the increasing Zeeman splitting ($\approx 0.3$ meV/T) of the QD spin doublet. The main Shiba peaks get strongly suppressed and they seem to eventually merge into the normal-state, Zeeman-split Kondo peaks through a non-trivial transition region (roughly between 120 and 180 mT). This large Zeeman splitting prevents the observation of a zero-bias Kondo peak when approaching $B^{\parallel}_c$. 

In conclusion, we have studied the transport properties of a spin-1/2 QD coupled to \textit{S} contacts. The ability to continuously tune $\Delta$ with an external magnetic field enabled us to investigate the transition from a normal-state Kondo to a superconducting-state Shiba ground state. We showed that the presence of a finite quasiparticle DOS within $\Delta$ can promote the formation of a zero-bias Kondo peak coexisting with superconductivity. The origin of this quasiparticle DOS remains to be clarified, also through a better understanding of the superconducting proximity effect in semiconductor NW structures (e.g., the role of disorder-induced pair breaking \cite{Plee}). 
Finally, we should like to emphasize that our results bear clear implications in the interpretation of sub-gap transport features in hybrid superconductor-semiconductor systems, especially 
in the presence of relatively high $B$ that can cause a significant suppression of $\Delta$. This is precisely the regime where Majorana fermions are expected to arise as zero-energy quasiparticle states. We argue that intra-gap quasiparticle states, manifesting through a sizable background conductance (as seen in Refs.~\onlinecite{[]MajoDelft, MajoXu, Weizmann}), could result in the screening of a local spin (or orbital) degeneracy leading to zero-bias anomalies that are not related to Majorana physics.  

This work was supported by the EU Marie Curie program and by the Agence Nationale de la Recherche. R. A. acknowledges support from the Spanish
Ministry of Science and Innovation through grant FIS2009-08744.

\bibliography{leezbp}

\begin{thebibliography}{35}%
\makeatletter
\providecommand \@ifxundefined [1]{%
 \@ifx{#1\undefined}
}%
\providecommand \@ifnum [1]{%
 \ifnum #1\expandafter \@firstoftwo
 \else \expandafter \@secondoftwo
 \fi
}%
\providecommand \@ifx [1]{%
 \ifx #1\expandafter \@firstoftwo
 \else \expandafter \@secondoftwo
 \fi
}%
\providecommand \natexlab [1]{#1}%
\providecommand \enquote  [1]{``#1''}%
\providecommand \bibnamefont  [1]{#1}%
\providecommand \bibfnamefont [1]{#1}%
\providecommand \citenamefont [1]{#1}%
\providecommand \href@noop [0]{\@secondoftwo}%
\providecommand \href [0]{\begingroup \@sanitize@url \@href}%
\providecommand \@href[1]{\@@startlink{#1}\@@href}%
\providecommand \@@href[1]{\endgroup#1\@@endlink}%
\providecommand \@sanitize@url [0]{\catcode `\\12\catcode `\$12\catcode
  `\&12\catcode `\#12\catcode `\^12\catcode `\_12\catcode `\%12\relax}%
\providecommand \@@startlink[1]{}%
\providecommand \@@endlink[0]{}%
\providecommand \url  [0]{\begingroup\@sanitize@url \@url }%
\providecommand \@url [1]{\endgroup\@href {#1}{\urlprefix }}%
\providecommand \urlprefix  [0]{URL }%
\providecommand \Eprint [0]{\href }%
\providecommand \doibase [0]{http://dx.doi.org/}%
\providecommand \selectlanguage [0]{\@gobble}%
\providecommand \bibinfo  [0]{\@secondoftwo}%
\providecommand \bibfield  [0]{\@secondoftwo}%
\providecommand \translation [1]{[#1]}%
\providecommand \BibitemOpen [0]{}%
\providecommand \bibitemStop [0]{}%
\providecommand \bibitemNoStop [0]{.\EOS\space}%
\providecommand \EOS [0]{\spacefactor3000\relax}%
\providecommand \BibitemShut  [1]{\csname bibitem#1\endcsname}%
\let\auto@bib@innerbib\@empty
\bibitem [{\citenamefont {Franceschi}\ \emph {et~al.}(2010)\citenamefont
  {Franceschi}, \citenamefont {Kouwenhoven}, \citenamefont {Schonenberger},\
  and\ \citenamefont {Wernsdorfer}}]{ReviewDF}%
  \BibitemOpen
  \bibfield  {author} {\bibinfo {author} {\bibfnamefont {S.~D.}\ \bibnamefont
  {Franceschi}}, \bibinfo {author} {\bibfnamefont {L.~P.}\ \bibnamefont
  {Kouwenhoven}}, \bibinfo {author} {\bibfnamefont {C.}~\bibnamefont
  {Schonenberger}}, \ and\ \bibinfo {author} {\bibfnamefont {W.}~\bibnamefont
  {Wernsdorfer}},\ }\href@noop {} {\bibfield  {journal} {\bibinfo  {journal}
  {Nature Nanotech.}\ }\textbf {\bibinfo {volume} {5}},\ \bibinfo {pages} {703}
  (\bibinfo {year} {2010})}\BibitemShut {NoStop}%
\bibitem [{\citenamefont {Lutchyn}\ \emph {et~al.}(2010)\citenamefont
  {Lutchyn}, \citenamefont {Sau},\ and\ \citenamefont {Sarma}}]{DasSarma}%
  \BibitemOpen
  \bibfield  {author} {\bibinfo {author} {\bibfnamefont {R.~M.}\ \bibnamefont
  {Lutchyn}}, \bibinfo {author} {\bibfnamefont {J.~D.}\ \bibnamefont {Sau}}, \
  and\ \bibinfo {author} {\bibfnamefont {S.~D.}\ \bibnamefont {Sarma}},\
  }\href@noop {} {\bibfield  {journal} {\bibinfo  {journal} {Phys.\ Rev.
  Lett.}\ }\textbf {\bibinfo {volume} {105}},\ \bibinfo {pages} {077001}
  (\bibinfo {year} {2010})}\BibitemShut {NoStop}%
\bibitem [{\citenamefont {Mourik}\ \emph {et~al.}(2012)\citenamefont {Mourik},
  \citenamefont {Zuo}, \citenamefont {Frolov}, \citenamefont {Plissard},
  \citenamefont {Bakkers},\ and\ \citenamefont {Kouwenhoven}}]{MajoDelft}%
  \BibitemOpen
  \bibfield  {author} {\bibinfo {author} {\bibfnamefont {V.}~\bibnamefont
  {Mourik}}, \bibinfo {author} {\bibfnamefont {K.}~\bibnamefont {Zuo}},
  \bibinfo {author} {\bibfnamefont {S.~M.}\ \bibnamefont {Frolov}}, \bibinfo
  {author} {\bibfnamefont {S.~R.}\ \bibnamefont {Plissard}}, \bibinfo {author}
  {\bibfnamefont {E.~P. A.~M.}\ \bibnamefont {Bakkers}}, \ and\ \bibinfo
  {author} {\bibfnamefont {L.~P.}\ \bibnamefont {Kouwenhoven}},\ }\href@noop {}
  {\bibfield  {journal} {\bibinfo  {journal} {Science}\ }\textbf {\bibinfo
  {volume} {336}},\ \bibinfo {pages} {1003} (\bibinfo {year}
  {2012})}\BibitemShut {NoStop}%
\bibitem [{\citenamefont {Deng}\ \emph {et~al.}(2012)\citenamefont {Deng},
  \citenamefont {Yu}, \citenamefont {Huang}, \citenamefont {Larsson},
  \citenamefont {Caroff},\ and\ \citenamefont {Xu}}]{MajoXu}%
  \BibitemOpen
  \bibfield  {author} {\bibinfo {author} {\bibfnamefont {M.~T.}\ \bibnamefont
  {Deng}}, \bibinfo {author} {\bibfnamefont {C.~L.}\ \bibnamefont {Yu}},
  \bibinfo {author} {\bibfnamefont {G.~Y.}\ \bibnamefont {Huang}}, \bibinfo
  {author} {\bibfnamefont {M.}~\bibnamefont {Larsson}}, \bibinfo {author}
  {\bibfnamefont {P.}~\bibnamefont {Caroff}}, \ and\ \bibinfo {author}
  {\bibfnamefont {H.~Q.}\ \bibnamefont {Xu}},\ }\href@noop {} {\bibfield
  {journal} {\bibinfo  {journal} {arXiv:1204.4130v1}\ } (\bibinfo {year}
  {2012})}\BibitemShut {NoStop}%
\bibitem [{\citenamefont {Das}\ \emph {et~al.}(2012{\natexlab{a}})\citenamefont
  {Das}, \citenamefont {Ronen}, \citenamefont {Most}, \citenamefont {Oreg},
  \citenamefont {Heiblum},\ and\ \citenamefont {Shtrikman}}]{Weizmann}%
  \BibitemOpen
  \bibfield  {author} {\bibinfo {author} {\bibfnamefont {A.}~\bibnamefont
  {Das}}, \bibinfo {author} {\bibfnamefont {Y.}~\bibnamefont {Ronen}}, \bibinfo
  {author} {\bibfnamefont {Y.}~\bibnamefont {Most}}, \bibinfo {author}
  {\bibfnamefont {Y.}~\bibnamefont {Oreg}}, \bibinfo {author} {\bibfnamefont
  {M.}~\bibnamefont {Heiblum}}, \ and\ \bibinfo {author} {\bibfnamefont
  {H.}~\bibnamefont {Shtrikman}},\ }\href@noop {} {\bibfield  {journal}
  {\bibinfo  {journal} {arXiv:1205.7073v1}\ } (\bibinfo {year}
  {2012}{\natexlab{a}})}\BibitemShut {NoStop}%
\bibitem [{\citenamefont {van Dam}\ \emph {et~al.}(2006)\citenamefont {van
  Dam}, \citenamefont {Nazarov}, \citenamefont {Bakkers}, \citenamefont
  {Franceschi},\ and\ \citenamefont {Kouwenhoven}}]{sdf2006}%
  \BibitemOpen
  \bibfield  {author} {\bibinfo {author} {\bibfnamefont {J.~A.}\ \bibnamefont
  {van Dam}}, \bibinfo {author} {\bibfnamefont {Y.~V.}\ \bibnamefont
  {Nazarov}}, \bibinfo {author} {\bibfnamefont {E.~P. A.~M.}\ \bibnamefont
  {Bakkers}}, \bibinfo {author} {\bibfnamefont {S.~D.}\ \bibnamefont
  {Franceschi}}, \ and\ \bibinfo {author} {\bibfnamefont {L.~P.}\ \bibnamefont
  {Kouwenhoven}},\ }\href@noop {} {\bibfield  {journal} {\bibinfo  {journal}
  {Nature}\ }\textbf {\bibinfo {volume} {442}},\ \bibinfo {pages} {667}
  (\bibinfo {year} {2006})}\BibitemShut {NoStop}%
\bibitem [{\citenamefont {Cleuziou}\ \emph {et~al.}(2006)\citenamefont
  {Cleuziou}, \citenamefont {Wernsdorfer}, \citenamefont {Bouchiat},
  \citenamefont {Ondarcuhu},\ and\ \citenamefont {Monthioux}}]{Cleuziou}%
  \BibitemOpen
  \bibfield  {author} {\bibinfo {author} {\bibfnamefont {J.}~\bibnamefont
  {Cleuziou}}, \bibinfo {author} {\bibfnamefont {W.}~\bibnamefont
  {Wernsdorfer}}, \bibinfo {author} {\bibfnamefont {V.}~\bibnamefont
  {Bouchiat}}, \bibinfo {author} {\bibfnamefont {T.}~\bibnamefont {Ondarcuhu}},
  \ and\ \bibinfo {author} {\bibfnamefont {M.}~\bibnamefont {Monthioux}},\
  }\href@noop {} {\bibfield  {journal} {\bibinfo  {journal} {Nature Nanotech.}\
  }\textbf {\bibinfo {volume} {1}},\ \bibinfo {pages} {53} (\bibinfo {year}
  {2006})}\BibitemShut {NoStop}%
\bibitem [{\citenamefont {Hofstetter}\ \emph {et~al.}(2009)\citenamefont
  {Hofstetter}, \citenamefont {Csonka}, \citenamefont {Nygard},\ and\
  \citenamefont {Schonenberger}}]{splitter01}%
  \BibitemOpen
  \bibfield  {author} {\bibinfo {author} {\bibfnamefont {L.}~\bibnamefont
  {Hofstetter}}, \bibinfo {author} {\bibfnamefont {S.}~\bibnamefont {Csonka}},
  \bibinfo {author} {\bibfnamefont {J.}~\bibnamefont {Nygard}}, \ and\ \bibinfo
  {author} {\bibfnamefont {C.}~\bibnamefont {Schonenberger}},\ }\href@noop {}
  {\bibfield  {journal} {\bibinfo  {journal} {Nature}\ }\textbf {\bibinfo
  {volume} {461}},\ \bibinfo {pages} {960} (\bibinfo {year}
  {2009})}\BibitemShut {NoStop}%
\bibitem [{\citenamefont {Herrmann}\ \emph {et~al.}(2010)\citenamefont
  {Herrmann}, \citenamefont {Portier}, \citenamefont {Roche}, \citenamefont
  {Yeyati}, \citenamefont {Kontos},\ and\ \citenamefont {Strunk}}]{splitter02}%
  \BibitemOpen
  \bibfield  {author} {\bibinfo {author} {\bibfnamefont {L.~G.}\ \bibnamefont
  {Herrmann}}, \bibinfo {author} {\bibfnamefont {F.}~\bibnamefont {Portier}},
  \bibinfo {author} {\bibfnamefont {P.}~\bibnamefont {Roche}}, \bibinfo
  {author} {\bibfnamefont {A.~L.}\ \bibnamefont {Yeyati}}, \bibinfo {author}
  {\bibfnamefont {T.}~\bibnamefont {Kontos}}, \ and\ \bibinfo {author}
  {\bibfnamefont {C.}~\bibnamefont {Strunk}},\ }\href@noop {} {\bibfield
  {journal} {\bibinfo  {journal} {Phys.\ Rev. Lett.}\ }\textbf {\bibinfo
  {volume} {104}},\ \bibinfo {pages} {026801} (\bibinfo {year}
  {2010})}\BibitemShut {NoStop}%
\bibitem [{\citenamefont {Das}\ \emph {et~al.}(2012{\natexlab{b}})\citenamefont
  {Das}, \citenamefont {Ronen}, \citenamefont {Heiblum}, \citenamefont
  {Mahalu}, \citenamefont {Kretinin},\ and\ \citenamefont
  {Shtrikman}}]{Dascooper}%
  \BibitemOpen
  \bibfield  {author} {\bibinfo {author} {\bibfnamefont {A.}~\bibnamefont
  {Das}}, \bibinfo {author} {\bibfnamefont {Y.}~\bibnamefont {Ronen}}, \bibinfo
  {author} {\bibfnamefont {M.}~\bibnamefont {Heiblum}}, \bibinfo {author}
  {\bibfnamefont {D.}~\bibnamefont {Mahalu}}, \bibinfo {author} {\bibfnamefont
  {A.~V.}\ \bibnamefont {Kretinin}}, \ and\ \bibinfo {author} {\bibfnamefont
  {H.}~\bibnamefont {Shtrikman}},\ }\href@noop {} {\bibfield  {journal}
  {\bibinfo  {journal} {arXiv:1205.2455v1}\ } (\bibinfo {year}
  {2012}{\natexlab{b}})}\BibitemShut {NoStop}%
\bibitem [{\citenamefont {Pillet}\ \emph {et~al.}(2010)\citenamefont {Pillet},
  \citenamefont {Quay}, \citenamefont {Morfin}, \citenamefont {Bena},
  \citenamefont {Yeyati},\ and\ \citenamefont {Joyez}}]{Pillet}%
  \BibitemOpen
  \bibfield  {author} {\bibinfo {author} {\bibfnamefont {J.~D.}\ \bibnamefont
  {Pillet}}, \bibinfo {author} {\bibfnamefont {C.~H.~L.}\ \bibnamefont {Quay}},
  \bibinfo {author} {\bibfnamefont {P.}~\bibnamefont {Morfin}}, \bibinfo
  {author} {\bibfnamefont {C.}~\bibnamefont {Bena}}, \bibinfo {author}
  {\bibfnamefont {A.~L.}\ \bibnamefont {Yeyati}}, \ and\ \bibinfo {author}
  {\bibfnamefont {P.}~\bibnamefont {Joyez}},\ }\href@noop {} {\bibfield
  {journal} {\bibinfo  {journal} {Nature Phys.}\ }\textbf {\bibinfo {volume}
  {6}},\ \bibinfo {pages} {965} (\bibinfo {year} {2010})}\BibitemShut {NoStop}%
\bibitem [{\citenamefont {Deacon}\ \emph {et~al.}(2010)\citenamefont {Deacon},
  \citenamefont {Tanaka}, \citenamefont {Oiwa}, \citenamefont {Sakano},
  \citenamefont {Yoshida}, \citenamefont {Shibata}, \citenamefont {Hirakawa},\
  and\ \citenamefont {Tarucha}}]{Deacon2010}%
  \BibitemOpen
  \bibfield  {author} {\bibinfo {author} {\bibfnamefont {R.~S.}\ \bibnamefont
  {Deacon}}, \bibinfo {author} {\bibfnamefont {Y.}~\bibnamefont {Tanaka}},
  \bibinfo {author} {\bibfnamefont {A.}~\bibnamefont {Oiwa}}, \bibinfo {author}
  {\bibfnamefont {R.}~\bibnamefont {Sakano}}, \bibinfo {author} {\bibfnamefont
  {K.}~\bibnamefont {Yoshida}}, \bibinfo {author} {\bibfnamefont
  {K.}~\bibnamefont {Shibata}}, \bibinfo {author} {\bibfnamefont
  {K.}~\bibnamefont {Hirakawa}}, \ and\ \bibinfo {author} {\bibfnamefont
  {S.}~\bibnamefont {Tarucha}},\ }\href@noop {} {\bibfield  {journal} {\bibinfo
   {journal} {Phys.\ Rev. Lett.}\ }\textbf {\bibinfo {volume} {104}},\ \bibinfo
  {pages} {076805} (\bibinfo {year} {2010})}\BibitemShut {NoStop}%
\bibitem [{\citenamefont {Meng}\ \emph {et~al.}(2009)\citenamefont {Meng},
  \citenamefont {Florens},\ and\ \citenamefont {Simon}}]{Simon}%
  \BibitemOpen
  \bibfield  {author} {\bibinfo {author} {\bibfnamefont {T.}~\bibnamefont
  {Meng}}, \bibinfo {author} {\bibfnamefont {S.}~\bibnamefont {Florens}}, \
  and\ \bibinfo {author} {\bibfnamefont {P.}~\bibnamefont {Simon}},\
  }\href@noop {} {\bibfield  {journal} {\bibinfo  {journal} {Phys. Rev. B}\
  }\textbf {\bibinfo {volume} {79}},\ \bibinfo {pages} {224521} (\bibinfo
  {year} {2009})}\BibitemShut {NoStop}%
\bibitem [{\citenamefont {Buitelaar}\ \emph {et~al.}(2002)\citenamefont
  {Buitelaar}, \citenamefont {Nussbaumer},\ and\ \citenamefont
  {Schonenberger}}]{Buitelaar}%
  \BibitemOpen
  \bibfield  {author} {\bibinfo {author} {\bibfnamefont {M.~R.}\ \bibnamefont
  {Buitelaar}}, \bibinfo {author} {\bibfnamefont {T.}~\bibnamefont
  {Nussbaumer}}, \ and\ \bibinfo {author} {\bibfnamefont {C.}~\bibnamefont
  {Schonenberger}},\ }\href@noop {} {\bibfield  {journal} {\bibinfo  {journal}
  {Phys.\ Rev. Lett.}\ }\textbf {\bibinfo {volume} {89}},\ \bibinfo {pages}
  {256801} (\bibinfo {year} {2002})}\BibitemShut {NoStop}%
\bibitem [{\citenamefont {Siano}\ and\ \citenamefont {Egger}(2004)}]{Siano}%
  \BibitemOpen
  \bibfield  {author} {\bibinfo {author} {\bibfnamefont {F.}~\bibnamefont
  {Siano}}\ and\ \bibinfo {author} {\bibfnamefont {R.}~\bibnamefont {Egger}},\
  }\href@noop {} {\bibfield  {journal} {\bibinfo  {journal} {Phys.\ Rev.
  Lett.}\ }\textbf {\bibinfo {volume} {93}},\ \bibinfo {pages} {047002}
  (\bibinfo {year} {2004})}\BibitemShut {NoStop}%
\bibitem [{\citenamefont {Choi}\ \emph {et~al.}(2004)\citenamefont {Choi},
  \citenamefont {Lee}, \citenamefont {Kang},\ and\ \citenamefont
  {Belzig}}]{Choi}%
  \BibitemOpen
  \bibfield  {author} {\bibinfo {author} {\bibfnamefont {M.~S.}\ \bibnamefont
  {Choi}}, \bibinfo {author} {\bibfnamefont {M.}~\bibnamefont {Lee}}, \bibinfo
  {author} {\bibfnamefont {K.}~\bibnamefont {Kang}}, \ and\ \bibinfo {author}
  {\bibfnamefont {W.}~\bibnamefont {Belzig}},\ }\href@noop {} {\bibfield
  {journal} {\bibinfo  {journal} {Phys.\ Rev. B}\ }\textbf {\bibinfo {volume}
  {70}},\ \bibinfo {pages} {020502(R)} (\bibinfo {year} {2004})}\BibitemShut
  {NoStop}%
\bibitem [{\citenamefont {Oguri}\ \emph {et~al.}(2004)\citenamefont {Oguri},
  \citenamefont {Tanaka},\ and\ \citenamefont {Hewson}}]{Oguri}%
  \BibitemOpen
  \bibfield  {author} {\bibinfo {author} {\bibfnamefont {A.}~\bibnamefont
  {Oguri}}, \bibinfo {author} {\bibfnamefont {Y.}~\bibnamefont {Tanaka}}, \
  and\ \bibinfo {author} {\bibfnamefont {A.~C.}\ \bibnamefont {Hewson}},\
  }\href@noop {} {\bibfield  {journal} {\bibinfo  {journal} {J. Phys. Soc.
  Japan}\ }\textbf {\bibinfo {volume} {73}},\ \bibinfo {pages} {2494} (\bibinfo
  {year} {2004})}\BibitemShut {NoStop}%
\bibitem [{\citenamefont {Buizert}\ \emph {et~al.}(2007)\citenamefont
  {Buizert}, \citenamefont {Oiwa}, \citenamefont {Shibata}, \citenamefont
  {Hirakawa},\ and\ \citenamefont {Tarucha}}]{Buizert}%
  \BibitemOpen
  \bibfield  {author} {\bibinfo {author} {\bibfnamefont {C.}~\bibnamefont
  {Buizert}}, \bibinfo {author} {\bibfnamefont {A.}~\bibnamefont {Oiwa}},
  \bibinfo {author} {\bibfnamefont {K.}~\bibnamefont {Shibata}}, \bibinfo
  {author} {\bibfnamefont {K.}~\bibnamefont {Hirakawa}}, \ and\ \bibinfo
  {author} {\bibfnamefont {S.}~\bibnamefont {Tarucha}},\ }\href@noop {}
  {\bibfield  {journal} {\bibinfo  {journal} {Phys.\ Rev. Lett.}\ }\textbf
  {\bibinfo {volume} {99}},\ \bibinfo {pages} {136806} (\bibinfo {year}
  {2007})}\BibitemShut {NoStop}%
\bibitem [{\citenamefont {Eichler}\ \emph {et~al.}(2007)\citenamefont
  {Eichler}, \citenamefont {Weiss}, \citenamefont {Oberholzer}, \citenamefont
  {Schonenberger}, \citenamefont {Yeyati}, \citenamefont {Cuevas},\ and\
  \citenamefont {Martin-Rodero}}]{Eichler}%
  \BibitemOpen
  \bibfield  {author} {\bibinfo {author} {\bibfnamefont {A.}~\bibnamefont
  {Eichler}}, \bibinfo {author} {\bibfnamefont {M.}~\bibnamefont {Weiss}},
  \bibinfo {author} {\bibfnamefont {S.}~\bibnamefont {Oberholzer}}, \bibinfo
  {author} {\bibfnamefont {C.}~\bibnamefont {Schonenberger}}, \bibinfo {author}
  {\bibfnamefont {A.~L.}\ \bibnamefont {Yeyati}}, \bibinfo {author}
  {\bibfnamefont {J.~C.}\ \bibnamefont {Cuevas}}, \ and\ \bibinfo {author}
  {\bibfnamefont {A.}~\bibnamefont {Martin-Rodero}},\ }\href@noop {} {\bibfield
   {journal} {\bibinfo  {journal} {Phys.\ Rev. Lett.}\ }\textbf {\bibinfo
  {volume} {99}},\ \bibinfo {pages} {126602} (\bibinfo {year}
  {2007})}\BibitemShut {NoStop}%
\bibitem [{\citenamefont {Sand-Jespersen}\ \emph {et~al.}(2007)\citenamefont
  {Sand-Jespersen}, \citenamefont {Paaske}, \citenamefont {Andersen},
  \citenamefont {Grove-Rasmussen}, \citenamefont {Jorgensen}, \citenamefont
  {Aagesen}, \citenamefont {Sorensen}, \citenamefont {Lindelof}, \citenamefont
  {Flensberg},\ and\ \citenamefont {Nygard}}]{Jesp}%
  \BibitemOpen
  \bibfield  {author} {\bibinfo {author} {\bibfnamefont {T.}~\bibnamefont
  {Sand-Jespersen}}, \bibinfo {author} {\bibfnamefont {J.}~\bibnamefont
  {Paaske}}, \bibinfo {author} {\bibfnamefont {B.~M.}\ \bibnamefont
  {Andersen}}, \bibinfo {author} {\bibfnamefont {K.}~\bibnamefont
  {Grove-Rasmussen}}, \bibinfo {author} {\bibfnamefont {H.~I.}\ \bibnamefont
  {Jorgensen}}, \bibinfo {author} {\bibfnamefont {M.}~\bibnamefont {Aagesen}},
  \bibinfo {author} {\bibfnamefont {C.~B.}\ \bibnamefont {Sorensen}}, \bibinfo
  {author} {\bibfnamefont {P.~E.}\ \bibnamefont {Lindelof}}, \bibinfo {author}
  {\bibfnamefont {K.}~\bibnamefont {Flensberg}}, \ and\ \bibinfo {author}
  {\bibfnamefont {J.}~\bibnamefont {Nygard}},\ }\href@noop {} {\bibfield
  {journal} {\bibinfo  {journal} {Phys.\ Rev. Lett.}\ }\textbf {\bibinfo
  {volume} {99}},\ \bibinfo {pages} {126603} (\bibinfo {year}
  {2007})}\BibitemShut {NoStop}%
\bibitem [{\citenamefont {Kanai}\ \emph {et~al.}(2010)\citenamefont {Kanai},
  \citenamefont {Deacon}, \citenamefont {Oiwa}, \citenamefont {Yoshida},
  \citenamefont {Shibata}, \citenamefont {Hirakawa},\ and\ \citenamefont
  {Tarucha}}]{Kanai}%
  \BibitemOpen
  \bibfield  {author} {\bibinfo {author} {\bibfnamefont {Y.}~\bibnamefont
  {Kanai}}, \bibinfo {author} {\bibfnamefont {R.~S.}\ \bibnamefont {Deacon}},
  \bibinfo {author} {\bibfnamefont {O.}~\bibnamefont {Oiwa}}, \bibinfo {author}
  {\bibfnamefont {K.}~\bibnamefont {Yoshida}}, \bibinfo {author} {\bibfnamefont
  {K.}~\bibnamefont {Shibata}}, \bibinfo {author} {\bibfnamefont
  {K.}~\bibnamefont {Hirakawa}}, \ and\ \bibinfo {author} {\bibfnamefont
  {S.}~\bibnamefont {Tarucha}},\ }\href@noop {} {\bibfield  {journal} {\bibinfo
   {journal} {Phys.\ Rev. B.}\ }\textbf {\bibinfo {volume} {82}},\ \bibinfo
  {pages} {054512} (\bibinfo {year} {2010})}\BibitemShut {NoStop}%
\bibitem [{\citenamefont {Grove-Rasmussen}\ \emph {et~al.}(2007)\citenamefont
  {Grove-Rasmussen}, \citenamefont {Jorgensen},\ and\ \citenamefont
  {Lindelof}}]{Kasper}%
  \BibitemOpen
  \bibfield  {author} {\bibinfo {author} {\bibfnamefont {K.}~\bibnamefont
  {Grove-Rasmussen}}, \bibinfo {author} {\bibfnamefont {H.}~\bibnamefont
  {Jorgensen}}, \ and\ \bibinfo {author} {\bibfnamefont {P.}~\bibnamefont
  {Lindelof}},\ }\href@noop {} {\bibfield  {journal} {\bibinfo  {journal} {New
  J. Phys.}\ }\textbf {\bibinfo {volume} {9}},\ \bibinfo {pages} {124}
  (\bibinfo {year} {2007})}\BibitemShut {NoStop}%
\bibitem [{\citenamefont {Eichler}\ \emph {et~al.}(2009)\citenamefont
  {Eichler}, \citenamefont {Deblock}, \citenamefont {Weiss}, \citenamefont
  {Karrasch}, \citenamefont {Meden},\ and\ \citenamefont
  {Schonenberger}}]{Eichler02}%
  \BibitemOpen
  \bibfield  {author} {\bibinfo {author} {\bibfnamefont {A.}~\bibnamefont
  {Eichler}}, \bibinfo {author} {\bibfnamefont {R.}~\bibnamefont {Deblock}},
  \bibinfo {author} {\bibfnamefont {M.}~\bibnamefont {Weiss}}, \bibinfo
  {author} {\bibfnamefont {C.}~\bibnamefont {Karrasch}}, \bibinfo {author}
  {\bibfnamefont {V.}~\bibnamefont {Meden}}, \ and\ \bibinfo {author}
  {\bibfnamefont {C.}~\bibnamefont {Schonenberger}},\ }\href@noop {} {\bibfield
   {journal} {\bibinfo  {journal} {Phys.\ Rev. B}\ }\textbf {\bibinfo {volume}
  {79}},\ \bibinfo {pages} {161407} (\bibinfo {year} {2009})}\BibitemShut
  {NoStop}%
\bibitem [{\citenamefont {Franke}\ \emph {et~al.}(2011)\citenamefont {Franke},
  \citenamefont {Schulze},\ and\ \citenamefont {Pascual}}]{stm}%
  \BibitemOpen
  \bibfield  {author} {\bibinfo {author} {\bibfnamefont {K.~J.}\ \bibnamefont
  {Franke}}, \bibinfo {author} {\bibfnamefont {G.}~\bibnamefont {Schulze}}, \
  and\ \bibinfo {author} {\bibfnamefont {J.~I.}\ \bibnamefont {Pascual}},\
  }\href@noop {} {\bibfield  {journal} {\bibinfo  {journal} {Science}\ }\textbf
  {\bibinfo {volume} {332}},\ \bibinfo {pages} {940} (\bibinfo {year}
  {2011})}\BibitemShut {NoStop}%
\bibitem [{\citenamefont {Glazman}\ and\ \citenamefont
  {Matveev}(1989)}]{Glazman}%
  \BibitemOpen
  \bibfield  {author} {\bibinfo {author} {\bibfnamefont {L.~I.}\ \bibnamefont
  {Glazman}}\ and\ \bibinfo {author} {\bibfnamefont {K.}~\bibnamefont
  {Matveev}},\ }\href@noop {} {\bibfield  {journal} {\bibinfo  {journal} {JETP
  Lett.}\ }\textbf {\bibinfo {volume} {49}},\ \bibinfo {pages} {659} (\bibinfo
  {year} {1989})}\BibitemShut {NoStop}%
\bibitem [{\citenamefont {Jiang}\ \emph {et~al.}(2007)\citenamefont {Jiang},
  \citenamefont {Xiong}, \citenamefont {Qian}, \citenamefont {Li},\ and\
  \citenamefont {Lieber}}]{Xiaocheng}%
  \BibitemOpen
  \bibfield  {author} {\bibinfo {author} {\bibfnamefont {X.}~\bibnamefont
  {Jiang}}, \bibinfo {author} {\bibfnamefont {Q.}~\bibnamefont {Xiong}},
  \bibinfo {author} {\bibfnamefont {F.}~\bibnamefont {Qian}}, \bibinfo {author}
  {\bibfnamefont {Y.}~\bibnamefont {Li}}, \ and\ \bibinfo {author}
  {\bibfnamefont {C.~M.}\ \bibnamefont {Lieber}},\ }\href@noop {} {\bibfield
  {journal} {\bibinfo  {journal} {Nano Lett.}\ }\textbf {\bibinfo {volume}
  {7}},\ \bibinfo {pages} {3214} (\bibinfo {year} {2007})}\BibitemShut
  {NoStop}%
\bibitem [{gam()}]{gamma}%
  \BibitemOpen
  \href@noop {} {\ }\bibinfo {note} {In our modelling we used $\gamma(B) =
  0.4(B/B_c)\Delta(B)$. This simple function captures the experimentally
  observed field-induced increase in the broadening of the BCS coherence peaks
  and in the quasiparticle DOS at the Fermi level (both linked to the
  $\gamma(B)/\Delta(B)$ ratio).}\BibitemShut {Stop}%
\bibitem [{\citenamefont {Koerting}\ \emph {et~al.}(2010)\citenamefont
  {Koerting}, \citenamefont {Andersen}, \citenamefont {Flensberg},\ and\
  \citenamefont {Paaske}}]{Paaske2010}%
  \BibitemOpen
  \bibfield  {author} {\bibinfo {author} {\bibfnamefont {V.}~\bibnamefont
  {Koerting}}, \bibinfo {author} {\bibfnamefont {B.~M.}\ \bibnamefont
  {Andersen}}, \bibinfo {author} {\bibfnamefont {K.}~\bibnamefont {Flensberg}},
  \ and\ \bibinfo {author} {\bibfnamefont {J.}~\bibnamefont {Paaske}},\
  }\href@noop {} {\bibfield  {journal} {\bibinfo  {journal} {Phys.\ Rev. B}\
  }\textbf {\bibinfo {volume} {82}},\ \bibinfo {pages} {245108} (\bibinfo
  {year} {2010})}\BibitemShut {NoStop}%
\bibitem [{\citenamefont {Andersen}\ \emph {et~al.}(2011)\citenamefont
  {Andersen}, \citenamefont {Flensberg}, \citenamefont {Koerting},\ and\
  \citenamefont {Paaske}}]{Paaske2011}%
  \BibitemOpen
  \bibfield  {author} {\bibinfo {author} {\bibfnamefont {B.~M.}\ \bibnamefont
  {Andersen}}, \bibinfo {author} {\bibfnamefont {K.}~\bibnamefont {Flensberg}},
  \bibinfo {author} {\bibfnamefont {V.}~\bibnamefont {Koerting}}, \ and\
  \bibinfo {author} {\bibfnamefont {J.}~\bibnamefont {Paaske}},\ }\href@noop {}
  {\bibfield  {journal} {\bibinfo  {journal} {Phys.\ Rev. Lett.}\ }\textbf
  {\bibinfo {volume} {107}},\ \bibinfo {pages} {256802} (\bibinfo {year}
  {2011})}\BibitemShut {NoStop}%
\bibitem [{\citenamefont {Yu}(1965)}]{Yu}%
  \BibitemOpen
  \bibfield  {author} {\bibinfo {author} {\bibfnamefont {L.}~\bibnamefont
  {Yu}},\ }\href@noop {} {\bibfield  {journal} {\bibinfo  {journal} {Acta Phys.
  Sin.}\ }\textbf {\bibinfo {volume} {21}},\ \bibinfo {pages} {75} (\bibinfo
  {year} {1965})}\BibitemShut {NoStop}%
\bibitem [{\citenamefont {Shiba}(1968)}]{Shiba}%
  \BibitemOpen
  \bibfield  {author} {\bibinfo {author} {\bibfnamefont {H.}~\bibnamefont
  {Shiba}},\ }\href@noop {} {\bibfield  {journal} {\bibinfo  {journal} {Prog.
  Theor. Phys.}\ }\textbf {\bibinfo {volume} {40}},\ \bibinfo {pages} {435}
  (\bibinfo {year} {1968})}\BibitemShut {NoStop}%
\bibitem [{\citenamefont {Rusinov}(1969)}]{Rusinov}%
  \BibitemOpen
  \bibfield  {author} {\bibinfo {author} {\bibfnamefont {A.~I.}\ \bibnamefont
  {Rusinov}},\ }\href@noop {} {\bibfield  {journal} {\bibinfo  {journal} {Zh.
  Eksp. Teor. Fiz.}\ }\textbf {\bibinfo {volume} {56}},\ \bibinfo {pages}
  {2047} (\bibinfo {year} {1969})},\ \bibinfo {note} {[Sov. Phys. JETP 29, 1101
  (1969)]}\BibitemShut {NoStop}%
\bibitem [{\citenamefont {Balatsky}\ \emph {et~al.}(2006)\citenamefont
  {Balatsky}, \citenamefont {Vekhter},\ and\ \citenamefont {Zhu}}]{Balatsky}%
  \BibitemOpen
  \bibfield  {author} {\bibinfo {author} {\bibfnamefont {A.~V.}\ \bibnamefont
  {Balatsky}}, \bibinfo {author} {\bibfnamefont {I.}~\bibnamefont {Vekhter}}, \
  and\ \bibinfo {author} {\bibfnamefont {J.~X.}\ \bibnamefont {Zhu}},\
  }\href@noop {} {\bibfield  {journal} {\bibinfo  {journal} {Rev. Mod. Phys.}\
  }\textbf {\bibinfo {volume} {78}},\ \bibinfo {pages} {373} (\bibinfo {year}
  {2006})}\BibitemShut {NoStop}%
\bibitem [{\citenamefont {Yazdani}\ \emph {et~al.}(1997)\citenamefont
  {Yazdani}, \citenamefont {Jones}, \citenamefont {Lutz}, \citenamefont
  {Crommie},\ and\ \citenamefont {Eigler}}]{Yazdani}%
  \BibitemOpen
  \bibfield  {author} {\bibinfo {author} {\bibfnamefont {A.}~\bibnamefont
  {Yazdani}}, \bibinfo {author} {\bibfnamefont {B.~A.}\ \bibnamefont {Jones}},
  \bibinfo {author} {\bibfnamefont {C.~P.}\ \bibnamefont {Lutz}}, \bibinfo
  {author} {\bibfnamefont {M.~F.}\ \bibnamefont {Crommie}}, \ and\ \bibinfo
  {author} {\bibfnamefont {D.~M.}\ \bibnamefont {Eigler}},\ }\href@noop {}
  {\bibfield  {journal} {\bibinfo  {journal} {Science}\ }\textbf {\bibinfo
  {volume} {275}},\ \bibinfo {pages} {1767} (\bibinfo {year}
  {1997})}\BibitemShut {NoStop}%
\bibitem [{\citenamefont {Liu}\ \emph {et~al.}(2012)\citenamefont {Liu},
  \citenamefont {Potter}, \citenamefont {Law},\ and\ \citenamefont
  {Lee}}]{Plee}%
  \BibitemOpen
  \bibfield  {author} {\bibinfo {author} {\bibfnamefont {J.}~\bibnamefont
  {Liu}}, \bibinfo {author} {\bibfnamefont {A.~C.}\ \bibnamefont {Potter}},
  \bibinfo {author} {\bibfnamefont {K.~T.}\ \bibnamefont {Law}}, \ and\
  \bibinfo {author} {\bibfnamefont {P.~A.}\ \bibnamefont {Lee}},\ }\href@noop
  {} {\bibfield  {journal} {\bibinfo  {journal} {arXiv:1206.1276v1}\ }
  (\bibinfo {year} {2012})}\BibitemShut {NoStop}%
\end{thebibliography}%

\end{document}


\title{Supplemental Material: Zero-bias anomaly in a nanowire quantum dot coupled to superconductors}

\author{Eduardo J. H. Lee$^{1}$}
\author{Xiaocheng Jiang$^{2}$}
\author{Ramon Aguado$^{3}$}
\author{Georgios Katsaros$^{1}$}
\author{Charles M. Lieber$^{2}$}
\author{Silvano De Franceschi$^{1}$}
\email{silvano.defranceschi@cea.fr}

\affiliation{$^{1}$SPSMS, CEA-INAC/UJF-Grenoble 1, 17 rue des Martyrs, 38054 Grenoble Cedex 9, France}
\affiliation{$^{2}$Harvard University, Department of Chemistry and Chemical Biology, Cambridge, MA, 02138, USA}
\affiliation{$^{3}$Instituto de Ciencia de Materiales de Madrid, ICMM-CSIC Cantoblanco, 28049 Madrid, Spain}

\maketitle
\subsection{Model}
The dot is described by an Anderson Hamiltonian 

\begin{eqnarray}
\label{su4vs2::eq:HD}
H_D = \sum_{\sigma=\uparrow,\downarrow}
\varepsilon_{\sigma} d_{\sigma}^\dag d_{\sigma}
+ Un_{\uparrow}n_{\downarrow} \,,
\end{eqnarray}
where $\varepsilon_{\sigma}$ is the single-particle energy level of the
localized state with spin $\sigma$, $d_{\sigma}^\dag$
($d_{\sigma}$) the fermion creation (annihilation) operator of the
state,
\begin{math}
n_{\sigma} = d_{\sigma}^\dag d_{\sigma}
\end{math}
the occupation, and $U$ the on-site Coulomb
interaction, which defines the
charging energy.
We focus on a regime where the charging energy is much bigger than other energy scales.  In
this regime the Hamiltonian in Eq.~(\ref{su4vs2::eq:HD}) suffices
to describe all relevant physics.

Kondo physics arises as a result of the interplay between the strong
correlation in the dot and the coupling of the localized electrons with
the electrons in conduction bands. In our case, these are described as two leads ($\alpha=L$ and $R$) which can be either normal or superconducting. In this later case, they are represented by a BCS hamiltonian of the form:
\begin{equation}
\label{su4vs2::eq:HC1}
H_\alpha =
\sum_{k_\alpha}\sum_\sigma\varepsilon_{k_\alpha,\sigma}\,
a_{k_\alpha\sigma}^\dag a_{k_\alpha\sigma} +\sum_{k_\alpha} \Delta
(a_{k_\alpha\uparrow}^\dag a_{-k_\alpha\downarrow}^\dag+h.c.)\,
\end{equation}
with $\Delta$ being the superconducting pairing gap.
Tunneling is described by the Hamiltonian
\begin{equation}
\label{su4vs2::eq:HT1}
H_T =
\sum_{k_\alpha\sigma}
\left(V_{k_\alpha\sigma}
  a_{k_\alpha\sigma}^\dag d_{\sigma}
  + h.c.\right) \,.
\end{equation}
The total Hamiltonian is then given by
\begin{math}
H = H_L + H_R + H_T + H_D
\end{math} \,. For simplicity, we ignore
the $k$- and $\sigma$-dependence of the tunneling amplitudes.
Therefore, we consider a simplified model with $V_{k_\alpha
\sigma}=V_{\alpha}/\sqrt{2}$ which defines the widths $ \Gamma_\alpha^N = \pi\rho_0 |V_\alpha|^2$,
where $\rho_0 $ is the (normal) density of states in the reservoirs. In the presence of superconductivity, the coupling to the reservoirs is modified due to the BCS density of states as
\begin{equation}
\Gamma_\alpha(E)=\Gamma_\alpha^NN_{s}(E,B)=\Gamma_\alpha^NRe\frac{|E|}{\sqrt{|E|^2-{\Delta(B)}^2}}.
\end{equation}
This, of course, corresponds to a clean BCS case. As discussed in the main text, it is clear from the experimental data in even diamonds that the clean BCS case is not a good description of the experiments.
Instead, a Dynes expression of the form
\begin{equation} 
N_{s}(E,\gamma,B)=Re[\frac{|E|+i\gamma(B)}{\sqrt{(|E|+i\gamma(B))^2-\Delta(B)^2}}],
\end{equation}
has to be used, where $\gamma(B)$ is a phenomenological broadening which takes into account a finite density of states inside the BCS gap.

\subsection{Non-crossing approximation method} 
Now we write the physical fermionic operator as a combination of a
pseudofermion and a boson operator as follows: $d_{\sigma}=b^\dagger
f_{\sigma}$ where $f_{,\sigma}$ is the pseudofermion which
annihilates one ''occupied state'' with spin $\sigma$,
and $b^\dagger$ is a boson operator which creates an ''empty
state''. We are interested in a limit where the Coulomb interaction is very
large such that we can safely take the limit of $U\rightarrow \infty$.  This
fact enforces the constraint
\begin{math}
\sum_{\sigma}f_{\sigma}^{\dag}f_{\sigma} + b^\dag b=1
\end{math},
that prevents the accommodation of two
electrons at the same time in QD level.  This constraint is treated with a Lagrange multiplier.

\begin{eqnarray}
\label{hamiltonian1}
H_\mathrm{SB}
&= &\sum_{k_L,\sigma}\varepsilon_{k_L,\sigma}\,
a_{k_L\sigma}^\dag a_{k_L\sigma} +\sum_{k_L} \Delta
(a_{k_L\uparrow}^\dag a_{-k_L\downarrow}^\dag+h.c.)+
\sum_{\sigma}\varepsilon_{\sigma}
f_{\sigma}^\dag f_{\sigma}
+ \frac{\overline{V}_L}{\sqrt{N}}\sum_{k_L,\sigma}
\, \left(c_{k_L,\sigma}^\dag b^\dag f_{\sigma} + h.c.\right) +(L\rightarrow R)\nonumber\\
&+&  \lambda \left(\sum_{\sigma} f_{\sigma}^\dag f_{\sigma}
  + b^\dag b - 1\right).
\end{eqnarray}
Notice that we have rescaled the tunneling amplitudes $
V_\alpha \to \overline{V}_\alpha\sqrt{N}$
according to the spirit of a $1/N$-expansion ($N$ is the total
degeneracy of the localized orbital).

Our next task is to solve this Hamiltonian, which is rather
complicated due to the presence of the three operators in the
tunneling part and the constrain. Moreover, we need to take into account superconductivity and non-equilibrium effects.  In order to do this we employ
the so-called Non-Crossing approximation (NCA) \cite{NCAneq1,NCAneq2,NCAneq3,ram03} generalized to the superconducting case \cite{Clerk,Kroha}. Without entering into much detail of the theory, we
just mention that the boson fields in Eq.~(\ref{hamiltonian1})
are treated as fluctuating operators such that both thermal and charge fluctuations are included in a
self-consistent manner. In particular, one has to derive self-consistent equations-of-motion
for the time-ordered double-time Green's function (sub-indexes are
omitted here):
\begin{eqnarray}
iG(t,t')&\equiv&\langle T_c f(t)f^\dagger(t')\rangle\nonumber\,,\\
iB(t,t')&\equiv&\langle T_c b(t)b^\dagger(t')\rangle,
\end{eqnarray}
or in terms of their analytic pieces:
\begin{eqnarray}
iG(t,t')&=&G^{>}(t,t')\theta(t-t')-G^{<}(t,t')\theta(t'-t)\,,\nonumber\\
iB(t,t')&=&B^{>}(t,t')\theta(t-t')+B^{<}(t,t')\theta(t'-t);
\end{eqnarray}

A rigorous and well established way to derive these
equations-of-motion was first introduced by Kadanoff and
Baym \cite{Kadan}, and has been related to other non-equilibrium
methods (like the Keldysh method) by Langreth, see
Ref.~\cite{lan76} for a review. In the paper, we just show numerical
results of the coupled set of integral NCA equations for our problem and refer the
interested reader to
Refs.~\cite{NCAneq1,NCAneq2,NCAneq3,ram03} for details. In particular, the density of states is given by
\begin{equation}
\rho(\omega)=-\frac{1}{\pi}\sum_{\sigma}
\mathrm{Im}[A^{r}_{\sigma}(\omega)],
\end{equation}
where $A^{r}_{\sigma}(\varepsilon)$ is the Fourier transform of
the retarded Green's function
$A^{r}_{\sigma}(t)=
G_{\sigma}^{r}(t)B^{<}(-t)-G^{<}_{\sigma}(t) B^{a}(-t)$. Note that this decoupling neglects vertex corrections and, as a result, the NCA fails in describing the low-energy Fermi-liquid regime. Nevertheless, the NCA has proven to give reliable results even at temperatures well below the Kondo temperature (of the order of $T=10^{-2}T_K$) \cite{Kroha2}.
Following Meir and Wingreen in Ref. \cite{Meir92a}, the current is given by:
\begin{equation}
I_{\alpha\in \{ L,R \}}=-\frac{2e}{h}\sum_{\sigma}\int d\epsilon
\Gamma_\alpha(\epsilon)[2Im A^r_{\sigma}(\epsilon)f_\alpha(\epsilon)
+A^<_{\sigma}(\epsilon)].\nonumber\\
\end{equation}
with $A^<_{\sigma}(\epsilon)$ the Fourier transform of 
$A^{<}_{\sigma}(t)=
iG_{\sigma}^{<}(t)[B^{r}(-t)-B^{a}(-t)]$ and $f_\alpha(\epsilon)=\frac{1}{1+e^{\frac{(\epsilon-\mu_\alpha)}{kT}}}$ the Fermi-Dirac function at each reservoir held at a chemical potential $\mu_\alpha$ such that the applied bias voltage is defined as $eV=\mu_R-\mu_L$. 

In practice, we self-consistently solve the NCA integral equations until good numerical convergence is reached. All $dI/dV$ calculations presented in the main text are done for finite temperatures $T=0.25T_K$ and increasing values of $\Delta$, roughly from $\Delta=0$ (bottom curve of Fig. 1c in the main text) to $\Delta=20T_K$ (top curve of Fig. 1c in the main text).

\subsection{"Soft" gaps in out-of-plane and in-plane magnetic fields}

\begin{figure}
\includegraphics[width=86mm]{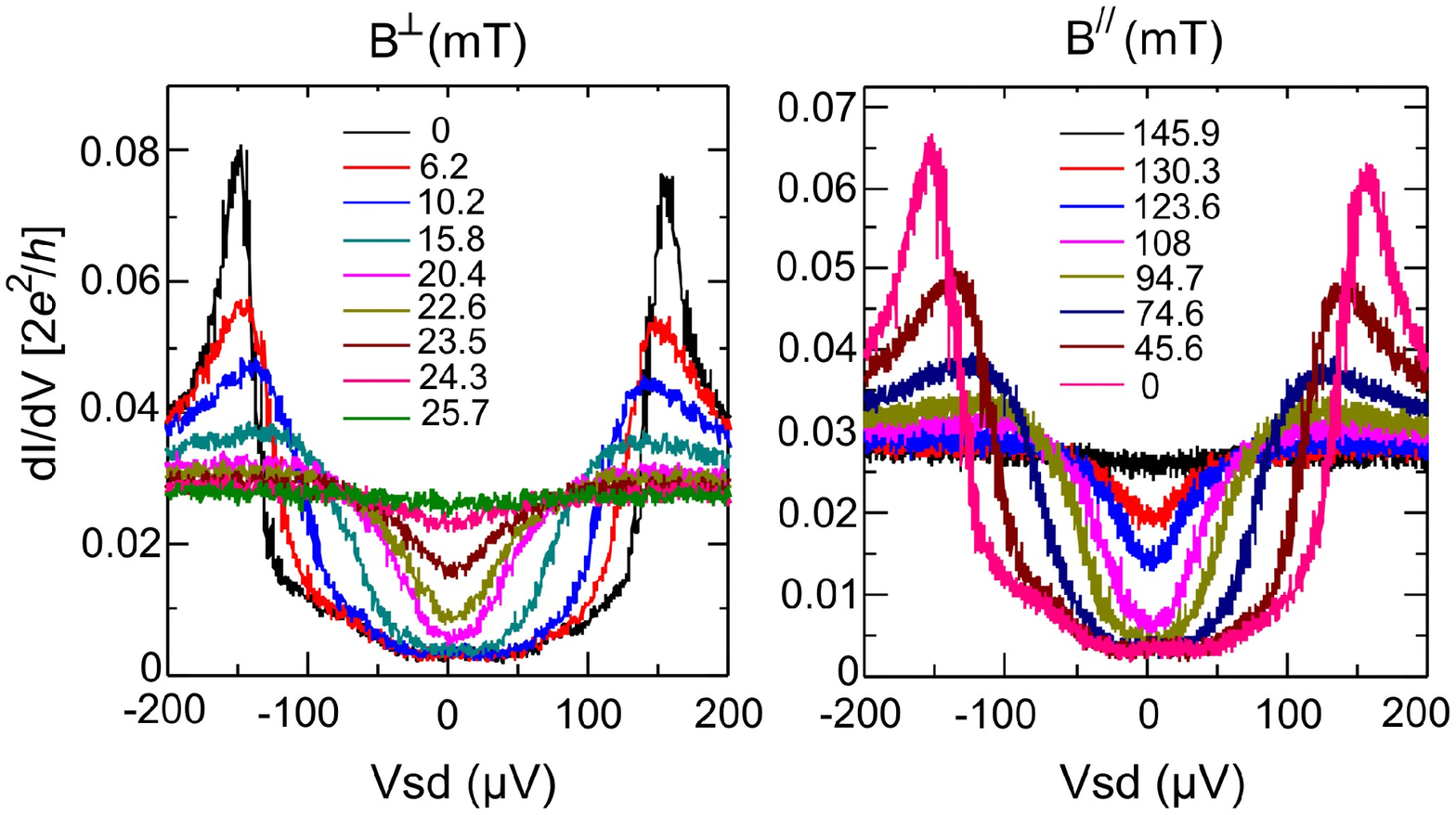}
\caption{\label{fig:epsart} Magnetic field dependence of the superconducting gap measured in an out-of-plane (left panel) or in an in-plane (right panel) configuration.}
\end{figure}

Here we present measurements in which the magnetic field dependence of the superconducting gap $\Delta$ was probed both in out-of-plane and in-plane configurations. For this purpose, we have performed measurements using S-QD-N devices, which are identical to the devices discussed in the main text, except for the fact that one of the S aluminium leads is replaced by a normal metal gold contact. The data shown in Fig. 1 was taken in an even valley, i.e. in the absence of Kondo correlations. As discussed in the main text, the transition from the normal state to the superconducting state is characterized by the opening of a "soft" gap, which is populated by intra-gap quasiparticle states. As $B$ decreases, the intra-gap DOS decreases, until a well-defined gap is obtained for low fields. The out-of-plane and in-plane field dependences are qualitatively identical, except for the different critical fields.

\subsection{On the origin of the "soft" gaps}

In the context of a more rigorous microscopic analysis in terms of Abrikosov-Gorkov (AG) theory \cite{AG60}, the density of states in the presence of a depairing mechanism can be written as \cite{Skalski64}
\begin{eqnarray}
N_s=\frac{u}{\sqrt{u^2-1}},
\end{eqnarray}
where $u$ is the complex solution of the equation:
\begin{eqnarray}
u\Delta(\gamma)=E+i\gamma\frac{u}{\sqrt{u^2-1}},
\end{eqnarray}
where $\gamma=\frac{\hbar}{\tau_{dp}}$ is the pair-breaking rate and $\Delta(\gamma)$ is a self-consistent order parameter. Obviously, in the limit $\gamma\rightarrow 0$ one recovers the ideal BCS case with $u=E/\Delta(0)$.
The precise microscopic expression of 
$\tau_{dp}$ depends on the nature of the pair-breaking mechanism.
In the case of thin superconducting films in a parallel magnetic field, the pair breaking reads \cite{Maki,TinkhamBook}:
\begin{equation}
\tau_{dp}^{-1}=\frac{v_F l}{18}(\frac{\pi d B_{||}}{\Phi_0})^2,
\end{equation}
with $v_F$, $l$, $d$ and $\Phi_0=\frac{hc}{2e}$, the Fermi velocity, the elastic mean free path, the film thickness and the flux quantum, respectively. In the perpendicular case, the corresponding expression reads 
\begin{equation}
\tau_{dp}^{-1}=\frac{v_F l}{3}\frac{\pi B_\perp}{\Phi_0}.
\end{equation}
Relevant for our analysis is the fact that for $\gamma>\Delta(\gamma)$ there is a finite density of states inside the gap.
Also interesting for our study is the limit $\Delta(\gamma)\rightarrow 0$ where $u\Delta(\gamma)\rightarrow E+i\gamma$. In this limit, near the gap closing, Eq. (1) becomes 
\begin{eqnarray}
N_s=\frac{E+i\gamma}{\sqrt{(E+i\gamma)^2-\Delta^2}},
\end{eqnarray}
and one can make contact with the phenomenological Dynes form used in our calculations.

Using Eq. (14), with a diffusion constant $D=v_Fl/3\approx 22.5 cm^2s^{-1}$, measured in 30 nm-thick Al thin films \cite{diffconst}, we estimate $\tau_{dp}^{-1}\approx 6.68 \times 10^{10} s^{-1}$ for a perpendicular magnetic field of $B_\perp = 20 mT$, which is close to $B_c^{\perp}$. This gives $\gamma^{(AG)}\approx 0.2\Delta_0$ (where $\Delta_0 \approx 155 \mu eV$). In our modelling we have used the relation $\gamma(B) = 0.4(B/B_c)\Delta(B)$ to describe the phenomenological Dynes broadning term. Considering a critical field $B_c^{\perp} \approx 23 mT$ (as experimentally observed in, e.g., diamond $\alpha$), this relation yields $\gamma(B)$ ranging from $\approx 0.15\Delta_0$ to 
$\approx 0.2\Delta_0$ in the interval $9.5 mT < B^{\perp} < 21 mT$, which is in consistent with the value estimated from the AG theory. Hence, field-induced pair breaking can indeed provide a reasonable explanation for the observed softening of the superconducting gap near $B_c$. 

It is noteworthy that our assumption that $\gamma(B)$ is proportional to $\Delta(B)$ becomes unphysical (hence incompatible with the AG theory) when $B$ is very close to $B_c$. However, as discussed above, it does describe well the $\gamma$ values expected from theory over a wide field range, including that relevant to our experimental observations. 

\subsection{Zeeman splitting of the zero-bias peak in the normal state}

To complement the data shown in the main text, we include here the perpendicular magnetic field dependence of $dI/dV$ at the center of diamond $\gamma$, measured up to 3 T (Fig. 2). Due to the width of the zero-bias Kondo peak, the splitting of the Kondo resonance induced by the Zeeman effect is not obvious for magnetic fields slightly higher than the critical field. From the splitting at higher fields, we estimate a $g$-factor of approximately 5.  

\begin{figure}
\includegraphics[width=86mm]{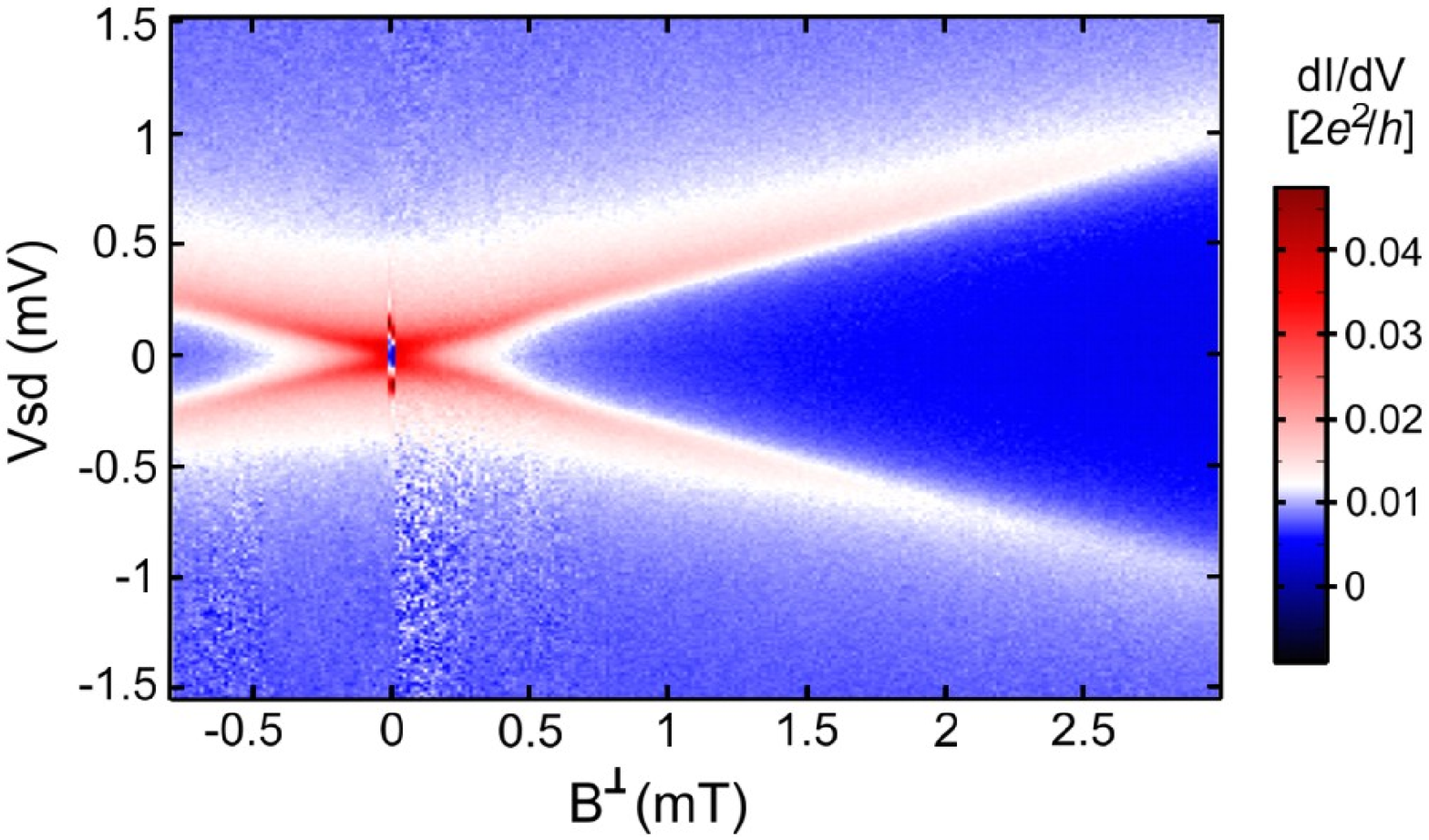}
\caption{\label{fig:epsart} Perpendicular magnetic field dependence of the $dI/dV$ measured in diamond $\gamma$.}
\end{figure}

\subsection{Temperature dependence of zero-bias Kondo peak}

We now discuss the temperature ($T$) dependence of the reported zero-bias Kondo anomalies in the regime of coexistence between the Kondo effect and superconductivity. We start by presenting data corresponding to Kondo resonance $\delta$ ($T_{K,\delta} \approx$ 0.35 K), which was not discussed in the main text. First, we verified that the overall magnetic field behavior of $\delta$ is qualitatively identical to that observed in diamonds $\alpha$, $\beta$ and $\gamma$ (described in the main text). Fig. 3a depicts the persistance and the narrowing of the zero-bias Kondo peak below the critical field $B_{c}^{\perp}$. Side peaks related to the superconducting gap are also visible. The temperature dependence measurements were taken at $B^{\perp}$ = 10 mT. Fig. 3b clearly demonstrates that the zero-bias Kondo peak and the finite-bias side peaks show distinct temperature dependences. Indeed, the zero-bias peak is strongly suppressed, while the side peaks are only weakly affected. 
This behavior relates to the fact that the height of the persisting Kondo anomaly follows the usual $T$-dependence for a normal state regime: $G(T)/G_{0}=[1/(1+(T/T_{K}^{'})^2]^s$, where $T_{K}^{'}=T_{K}^{*}/\sqrt{2^{\frac{1}{s}}-1}$, $s$ = 0.22 and $T_{K}^{*}$ is the effective Kondo temperature \cite{KondoT}. This implies that a 40$\%$ conductance drop is expected for $T$ = 0.2 K, assuming $T_{K,\delta} \approx$ 0.35 K (in fact $T_{K}^{*}$ is expected to be lower than the normal state $T_{K,\delta} \approx$ 0.35 K, as deduced from the narrowing of the zero-bias peak). By its turn,the superconducting gap at $B$=0, $\Delta_{0} \approx$ 150 $\mu$eV, is only reduced in $\approx$ 10$\%$ at $T$ = 0.2 K. 
As a consequence, the position of the side peaks in Fig. 3b are only marginally affected by $T$. 

\begin{figure}
\includegraphics[width=86mm]{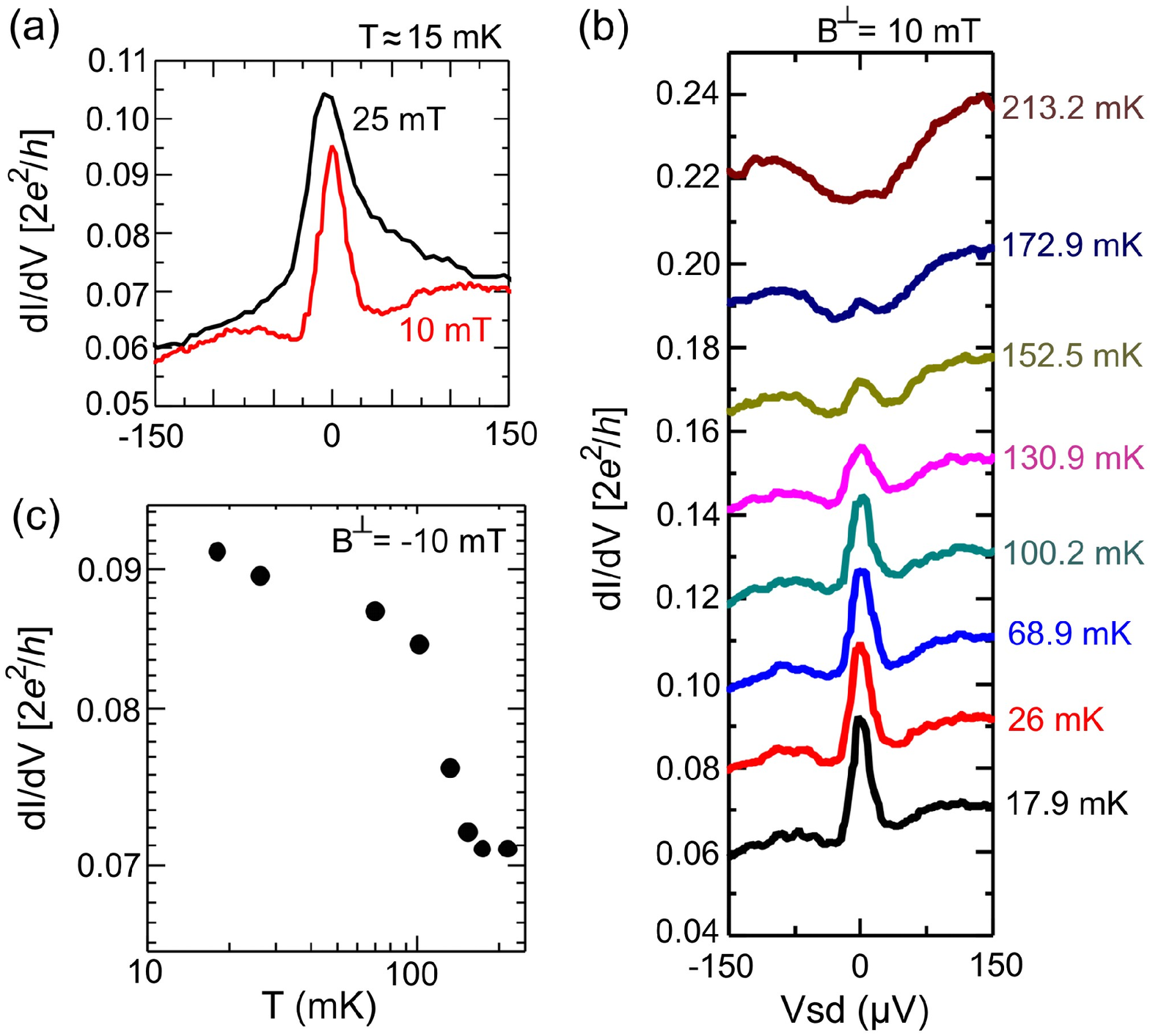}
\caption{\label{fig:epsart} (a) Persistance and narrowing of the Kondo peak below $B_{c}$. The side peaks observed in the red trace are related to the superconducting gap (see main text). (b) Temperature dependence of the persisting zero-bias Kondo anomaly. (c) $dI/dV$ of the zero-bias peak plotted as a function of the temperature.}
\end{figure}

\begin{figure}
\includegraphics[width=86mm]{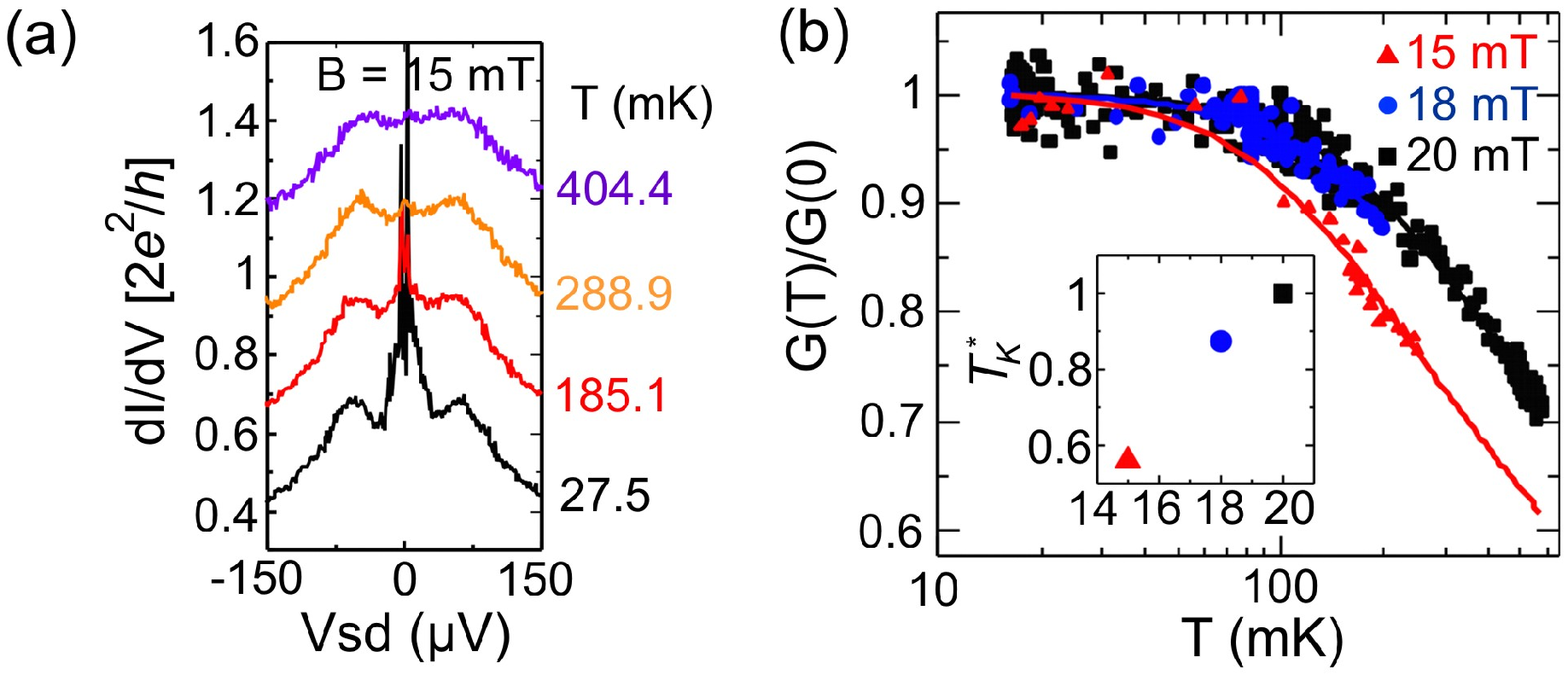}
\caption{\label{fig:epsart} (a) Evolution of the $dI/dV$ measured at $B^{\perp} \approx$ 15 mT, as a function of the temperature ($T$). The zero-bias Kondo peak and the gap-related side peaks show distinct $T$ dependences. The former is strongly suppressed with increasing $T$. (b) Temperature dependence of the normalized conductance of the zero-bias Kondo peak measured either above (squares) or below (circles and triangles) $B_{c}^{\perp}$. $G(0)$ denotes the peak height measured at base temperature. The inset reveals that the effective Kondo temperature $T_{K}^{*}$ decreases with increasing $\Delta$.}
\end{figure}

Temperature dependence measurements performed in diamond $\beta$ are shown in Fig. 4. The observed behavior is qualitatively identical to that of diamond $\delta$, i.e., the zero-bias Kondo peak is strongly suppressed by increasing $T$, while the side peaks related to $\Delta$ are only weakly affected. In addition, we show that $T_{K}^{*}$, obtained by fitting the above mentioned Kondo $T$ dependence to the experimental data, decreases with decreasing $B^{\perp}$ (inset of Fig. 4b), as expected from the reduced intra-gap quasiparticle DOS.